\documentclass[twoside]{article}

\newcommand{\bcA}{\mathbf{\cal A}}

\newcommand{\bbI}{{\mathbb I}}
\newcommand{\bbM}{{\mathbb M}}
\newcommand{\bbC}{{\mathbb C}}
\newcommand{\bbB}{{\mathbb B}}
\newcommand{\bbG}{{\mathbb G}}
\newcommand{\bbJ}{{\mathbb J}}

\newcommand{\bcG}{\mathbf{\cal G}}

\newcommand{\bde}{\begin{description}}

\newcommand{\ben}{\begin{enumerate}}
\newcommand{\beq}{\begin{eqnarray}}
\newcommand{\bet}{\boldsymbol{\eta}}
\newcommand{\beqn}{\begin{eqnarray*}}

\newcommand{\bmg}{{\mathbf g}}
\newcommand{\bmw}{{\mathbf w}}

\newcommand{\bmh}{{\mathbf h}}

\newcommand{\bqu}{\begin{quote}}

\newcommand{\btheta}{{\boldsymbol{\theta}}}

\newcommand{\bvb}{\begin{verbatim}}

\newcommand{\bxi}{\boldsymbol{\xi}}

\newcommand{\cf}{{\it cf.}~}

\newcommand{\ea}{\end{eqnarray}}

\newcommand{\ede}{\end{description}}
\newcommand{\een}{\end{enumerate}}
\newcommand{\eeq}{\end{eqnarray}}
\newcommand{\eeqn}{\end{eqnarray*}}
\newcommand{\eg}{{\it e.g.},~}
\newcommand{\etal}{et al.\ }

\newcommand{\equ}{\end{quote}}
\renewcommand{\etal}{{\it et al.}~}
\newcommand{\evb}{\end{verbatim}}

\newcommand{\goto}{\rightarrow}

\newcommand{\ie}{{\it i.e.},~}

\newcommand{\<}{\langle}

\newcommand{\oh }{\frac{1}{2}}

\newcommand{\via}{{{\it via}}}

\newcommand{\pmbeg}{\begin{pmatrix}}
\newcommand{\pmend}{\end{pmatrix}}

\renewcommand{\>}{\rangle}

\usepackage{amsmath} \usepackage{amssymb}

\newif\ifpdf \ifx\pdfoutput\undefined
\pdffalse 
\else
\pdfoutput=1 
\pdftrue \fi

\ifpdf\usepackage[pdftex]{graphicx} \else \usepackage{graphicx} \fi
\ifpdf\DeclareGraphicsExtensions{.pdf, .jpg, .tif} \else
\DeclareGraphicsExtensions{.eps, .ps} \fi

\ifpdf \usepackage[pdftex]{hyperref} \else
\usepackage[hypertex]{hyperref} \fi
 
\hypersetup{pdfsubject={Preprint NSF-ITP-02-47}, pdftitle={Process
    pathway inference via time series analysis}, pdfauthor={Chris
    Wiggins, Columbia University and Ilya Nemenman, UCSB},
  pdfkeywords={genomics, pathways, gene expression regulation,
    Bayesian statistic, auto-regressive models}, }

\begin{document}

\title{Process Pathway Inference \\via Time Series Analysis}

\author{
  Chris Wiggins$^{1,2}$ and Ilya Nemenman$^2$\\
  $^1$Department of Applied Physics and Applied Mathematics\\
  and Center for Computational Biology and Bioinformatics (C2B2)\\
  Columbia University, New York NY 10027\\
  $^2$Kavli Institute for Theoretical Physics\\
  University of California, Santa Barbara, CA 93106} \maketitle
\begin{abstract}
  Motivated by recent experimental developments in functional
  genomics, we construct and test a numerical technique for inferring
  {\it process pathways}, in which one process calls another process,
  from time series data.  We validate using a case in which data are
  readily available and formulate an extension, appropriate for
  genetic regulatory networks, which exploits Bayesian inference and
  in which the present--day undersampling is compensated for by prior
  understanding of genetic regulation.
\end{abstract}

Preprint number: NSF-ITP-02-47

\vskip .5 cm

\section{Motivation}

The last decade has witnessed stunning advances in experimental
biology, particularly in the fields of neuroscience and genomics,
which have made possible `data--driven' biological investigations.  As
examples, the quantitative revolution of genomics has provided
terabytes of transcriptome data; and neuroscientists routinely record
for hours or even days from multiple neurons simultaneously.  This
transformation stands as a challenge to theorists who hope to advance
understanding by making connection between experiment and first
principles models.\footnote{The mathematization of such models are
  referred to below as the `microscopic equations'; consider for
  example those of fluid dynamics which govern, yet certainly fail to
  encapsulate, such phenomena as turbulence and the tumbling of a
  falling leaf.}

In genomics, for example, we are presented with the expression levels
of thousands of genes, but our ability to model is limited not only
quantitatively, in that there are myriad unknown rate constants and
binding parameters, but qualitatively, in that a sizable fraction of
proteins and genes remain of uncharacterized function \cite{Stormo}.
Similarly, in neuroscience, we can model patches of cellular
membranes, synapses, and (at least electro--physiologically) entire
cells \cite{abbott-01}. However, this modeling hinges on numerous
unknown parameters, and even if we can perform massive computations
involved in the study of even rather small biological neural networks,
the sensitivity to these parameters still makes the whole approach
intractable.  The astronomical amounts of experimental data are
troubling computationally, but even more immediate problems are the
lack of reductive descriptions of the underlying phenomena and
undersampling --- an inability of the data to determine the (slightly
smaller) astronomical number of important microscopic parameters
appearing in theoretical models.

Presented with such an imbalance, it is important to distinguish among
the possible questions we can ask as well as the possible tools at our
disposal for answering them. That is, one can ask {\it what} the
system is doing (a nontrivial question when the language for discovery
is an astronomical number of unorganized data) before one asks {\it
  how} it is doing what it does. The latter involves building models
of some microscopic fidelity. The former may be answered without
reference to microscopics by a model--independent, data--driven
phenomenological approach.

A useful historical analogy is that of particle physics of the late
1950's, in which an explosion of data from accelerators was equally
daunting and similarly irreducible. At that time physicists were not
yet asking the {\it how} questions (cross sections, isospin
multiplets, etc.)  but were instead carefully, statistically,
inferring the presence or absence of features in the data; for
example, exploiting prior knowledge of quantum mechanics to constrain
reasonable shapes for peaks in the data (hallmarks of newly discovered
particles --- the `resonances').

In this analogy, neuroscience is still dealing with the existence of
peaks. Indeed, only recently (see Ref.~\cite{spikes} and further works
by the same authors) it has become clear that precise timings of
single spikes are very important for understanding the neural code.
This is a basic, objective, model--independent observation, and it is
not surprising that Shannon's information theory, which was
specifically designed with these types of questions in mind, turned
out to be extremely useful.

In genomics, however, the quantities of interest are easier to
identify: gene expressions are largely governed by the underlying
regulation networks. Now is the time to attempt to infer these
networks --- still the {\it what} question --- which corresponds to
inferring the peaks in the data.  This is a requisite step before
classifying the possible networks and explaining the classification
rules --- an answer to {\it how} the system does what it does.  Trying
to answer {\it how} before carefully exploring {\it what} might
ultimately produce many epi--manipulations of the data, but little
significant understanding.

Said otherwise, presented with data describing natural phenomena, one
should form a phenomenology of experimental results, then inferences
from the data in light of this phenomenology, and finally microscopic
models. Genomics is currently at the penultimate step, and, armed with
careful {\it informatics}, here meaning the incorporation of data with
prior knowledge in the absence of detailed models, we hope to reduce
the data into a representation which allows description, prediction,
and ultimately control.

An example of data reduction convenient for representation is
cataloging of the regulatory networks.  However, such cataloging is
not a model independent task: at the very least, our microscopic model
includes the existence of the networks.  Further, even if we are only
interested in a network's connection diagram and do not care much
about the exact details of the connections, identification of the
network still involves determination of many parameters. Thus it is
not clear that information--theoretic approaches will be of great use.
However, it is plausible that our intuition of how the underlying
microscopic dynamics translates into macroscopic probabilistic models
may play a big role. The main purpose of this paper is to show that
this intuition, appropriately mathematized in a principled way as the
{\it a priori} knowledge --- the priors in Bayesian statistics --- may
be self--consistently incorporated into a macroscopic probabilistic
model of process pathways without detailed, sophisticated modeling of
microscopic dynamics. We will first show this on a simple synthetic
example, and then suggest some extensions of the ideas with an eye
towards genetic regulatory networks.

\section{Functional genomics}
\label{funcgene}

As mentioned, the motivating problem here is time series informatics
applied to functional genomics.\footnote{It is important here to
  differentiate functional genomics, or `post--genomics', from
  sequencing genomics. The latter is the set of techniques associated
  with obtaining the genetic sequence of an organism. The former is
  the set of techniques which try to put this information to use.}  We
therefore briefly review genetics and characterize the relevant
experiments, the data from which will be used in inferring the
underlying connectivities and possibly control.

\subsection{A brief review of genetics}
The central goal of functional genomics is the understanding of the
interactions among distinct parts of the genome --- the {\it genes}.
Each gene consists of long words composed of thousands of coding base
pairs of DNA which are then transcribed into mRNA, which is then
translated into protein. Many of these proteins, called {\it
  transcription factors}, then regulate the rate of production of mRNA
transcribed either by their own genes or other genes.  The working of
all the genes thus forms a {\it genetic regulatory network}, and may
be thought of as a dynamical system.  Inputs include elements of the
physical world which affect the activity of the transcription factors,
and outputs may be considered as the concentrations of the translated
proteins or, at a deeper level, the transcribed mRNA. While the
proteins are ultimately responsible for cellular function, the mRNA
are more easily experimentally measured via {\it DNA microarrays}.

\subsection{A brief review of DNA microarrays}

Only recently has it become possible to probe the expression of a
number of genes comparable to the total number of genes in the entire
genome of an organism via microarrays of nucleic acids, commonly known
as `DNA chips.' The most common application of such a chip is to
monitor simultaneously the expression of thousands of genes by
detecting hybridization of nucleic acid originating from a biological
sample to target nucleic acids lying on the chip. One can then probe,
for example, the differences in gene expression between cancerous and
non--cancerous cells of the same specialization
\cite{CancerGenomics0,CancerGenomics1,CancerGenomics2,CancerGenomics3,CancerGenomics4,CancerGenomics5,CancerGenomics6},
or the expression of different genes as a function of the phase of the
cell cycle \cite{CellCycleGenomics1,CellCycleGenomics2}, or of the
response of cells to chemical or physical perturbation. The two latter
types of experiments produce time series of gene expressions, and they
will be the focal point of our discussion from now on.


The first DNA microarrays were made by Affymetrix in the early 1990s
\cite{Affy1,Affy2,Affy3}.  In this technique, DNA oligonucleotides are
attached to a surface in a specific spatial pattern, directed by
optically activated chemical synthesis.  One can build an arbitrary
oligonucleotide sequence in a small area (approximately 20 microns per
target) on the surface.  However, the initial setup cost of creating
the chip makes the technique infeasible for any application for which
less than several hundred masks will be created (and sold). Individual
researchers are completely without flexibility to change the chip to
fit a particular area of investigation.

Functional genomics further benefited from a second technology in
1996, when Pat Brown's lab at Stanford introduced the spot chip
\cite{POB1,POB2}. This highly customizable technique exploits robotic
deposition of drops on a microscope slide.  The automation makes
creating new and different slides a simple operation. Moreover, one
can create typically 120 slides at a time.  Individual researchers can
thus design custom experiments, placing genes at locations or
redundancies of their choosing.  The gene fragments used in the
spotting technique are hundreds of base pairs in length, and therefore
less sensitive to single base pair mismatches.


\section{Methods}

\subsection{Chemical network reconstruction} 

Genomic data is certainly not the first dynamic data for which reverse
engineering of the interaction network has been attempted.  A similar
problem has been faced historically in chemistry, in which one would
like to infer the underlying reactions responsible for observed data.
Such reaction networks typically are sparse, that is, the typical
connectivity is far less than the total number of chemical species.
This is also true in genomics, where one gene typically interacts with
no more than a few dozen others \cite{friedman00using}.

To highlight the parallels, one may state the question as follows:
armed with sufficient temporal data taken from a number of interacting
reagents (here, chemicals), is it possible to infer the circuit
diagram?  One possible strategy was proposed in 1995 \cite{Arkin95},
tested first on simulated data, and later on the glycolytic pathway
\cite{Arkin97}, and recently refined in light of ideas from
information theory \cite{Arkin01}. However, this strategy has yet to
be successfully applied to any reactions which were not known by the
authors beforehand, nor subjected to a `blind' test, as in the annual
CASP test among the protein folding community.\footnote{Critical
  Assessment of techniques for protein Structure Prediction;
  \hyperref{http://predictioncenter.llnl.gov/}{}{}{http://predictioncenter.llnl.gov/}.}
In addition, unlike in chemical kinetics, where the data is produced
by moderately nonlinear and rapidly interrogated dynamical systems,
genomic datasets are highly undersampled and are more like a set of
fuzzy logic gates, or leaky boolean circuits. Thus, successful
application of techniques inspired by chemical networks to genomic
data is, at best, doubtful.

We may nonetheless attempt reverse engineering of regulatory networks
with a similar philosophy, in that, rather than trying to fit to a
precise microscopic model, we attempt to parameterize a minimal
phenomenological model and infer macroscopic parameters from it.

\subsection{Synthetic network reconstruction}

In order to test any new attempt to infer connectivity from dynamics,
it is useful to study a system which is qualitatively similar, \eg
which demonstrates degrees of freedom which turn on and off other
degrees of freedom in a near--complete or `fuzzy logic gate' way and
for which connectivity is sparse; yet for which data are readily
available and obvious to interpret. To that end, we collected data
from a multiuser UNIX machine, recording the relative CPU usage of all
processes (the analogue of mRNA concentration), user ID, and process
name, as a function of time.  This can be done in an automated way via
\texttt{bash} script,\footnote{(GNU) Bourne--Again SHell} and the
results analyzed via MATLAB.

\begin{figure}[t]
  \centerline{\includegraphics[width=5in]{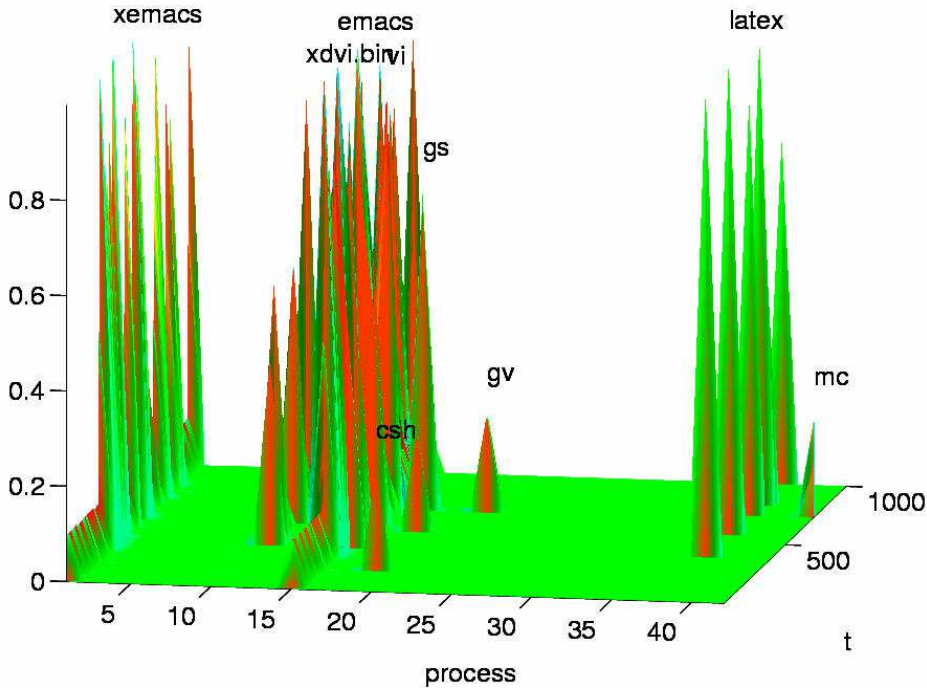}}
  \caption{CPU percentage for all the processes called by a particular
    user during the observation (approximately $10^3$ seconds).
    Processes are numbered according to their relative frequency over
    {\it all} users during the observation time.}
\label{x_vs_t}
\end{figure}

A typical time course, automatically labeled via MATLAB, is shown in
Fig.~\ref{x_vs_t} for a particular (anonymous) user at a large
department of applied mathematics.
One axis is the job `number,' ordered by frequency of occurrence over
all users; the other axis is time (roughly in seconds).  The height
(and color) indicates CPU usage.

\subsection{Modeling synthetic data}
\label{syndatmod}

We begin with the minimal probabilistic model of process pathways, in
which the strength of each process at subsequent time steps is
linearly determined by the strength of all current processes. Similar
linear models have been used with some success in understanding
genomic data, including
clustering via dynamics \cite{Ramoni}.  The most general model in
discrete time is the $AR(p)$ model, in which we include the
possibility that the state now is a function of the $p$ previous
observations of the system. We do not yet include the possibility of
hidden degrees of freedom.  Mathematically, we may pose the model as
\begin{eqnarray}
  \bmg_{t} &=& \bmw_0+\bbM_1 \bmg_{t-1}+\bbM_2\bmg_{t-2}+
  \ldots + \bbM_p\bmg_{t-p}+\bxi_t
  \label{synmod}\\
  \<\xi_{i,t}\xi_{j,t'}\>&=& C^{ij} \delta_{t,t'}, 
  \label{noisevar}
\end{eqnarray} 
where the degree of freedom at observation $t$ is $\bmg_t$, the
transition matrices are $\bbM_j$, and the noise correlation $\bbC$ is
as--yet undetermined.

One may fit for the most probable transition matrices $\bbM_{j}$ as
well as the offset $\bmw_0$ and the noise correlation matrix $\bbC$,
using, for example, the standard Schwartz's Bayesian Information
Criterion \cite{schwartz} to determine the most likely value of
$p$.\footnote{See Sec.~\ref{ss:bayes} for a brief discussion of this
  model selection technique.}  Excellent numerical techniques and
general purpose libraries have been designed for solving this problem
\cite{Tapio}.

Note that we {\em do not claim} the actual interactions among the
processes are linear, as Eq.~(\ref{synmod}) seems to imply. Indeed, as
stated above, the exact values of the transition matrices $\bbM_j$ are
of very little interest to us, and we are only interested in the
topological features of the network. It is reasonable to expect that
the absence of a connection between two processes will be fit well by
a zero in the corresponding transition matrix element, while the
presence of a connection of any type will result in its nonzero value.
The mismatch between the linear form of Eq.~(\ref{synmod}) and the
actual dynamics will manifest itself in a large variance $C^{ij}$.
However, if we are not interested in the exact values of $\bbM_j$,
this should not adversely effect our determination of connections.

\section{Results}
\label{ss:results}

\begin{figure}[t]
  \centerline{\includegraphics[width=5in] {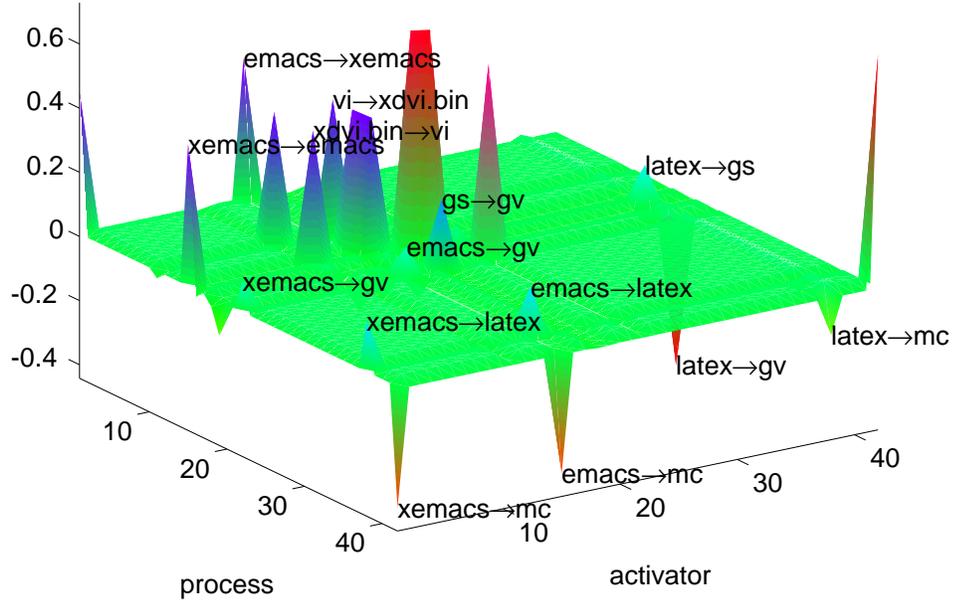}}
  \caption{The transition state matrix $\bbM_1$ resulting from the 
    data in Fig. \ref{x_vs_t}.}
  \label{Afig}
\end{figure}

Fitting the observed CPU usages to the transition state model, one
finds the most probable $p$ value is $1$, indicating a lack of inertia
in the system; the resulting transition matrix $\bbM_1$ is plotted in
Fig.~\ref{Afig}. The noise covariance matrix was quite small, despite
the naivet\'e of the model.

Causal connections between jobs are labeled with `$\goto$', \eg
`\texttt{emacs}$\goto$\texttt{latex}' or `\texttt{emacs} drives
\texttt{latex}.'  We highlight several remarkable features: \ben
  \item Processes familiar to anyone who has used the typesetting
software \LaTeX will be readily apparent: one edits a file (\eg in
\texttt{emacs}, \texttt{xemacs}, or \texttt{vi}), then compiles with
\texttt{latex}, and views the result in \texttt{xdvi} and finally
ghostscript (`\texttt{gs}'). Similarly one observes \texttt{emacs}
drives \texttt{latex}, \texttt{latex} drives \texttt{gs}, etc.
  \item Note that the matrix is not symmetric: one axis describes
processes which `activate' other processes; the second describes which
process is acted on. For example, \texttt{latex} drives ghostscript
but ghostscript does not drive \texttt{latex}.
  \item The transition matrix shows `upregulation' as well as
`downregulation': some processes discourage other processes at later
times.
  \item Diagonal elements have not been labeled as they simply
describe the likelihood the process will continue on to the next time
step.
  \item Note also how the transition matrix correctly infers the
highly sparse connectivity of these disparate jobs. The vast majority
of elements are $0$, as they should be, since processes are not
influenced by those called by other users.  \een

\begin{figure}[t]
  \centerline{\includegraphics[width=5in] {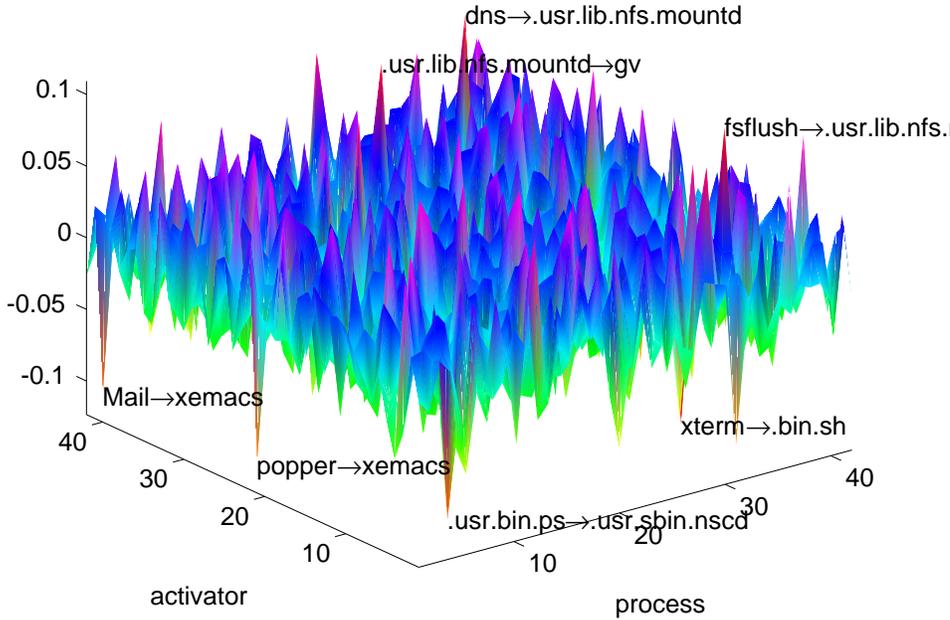}}
  \caption{The transition state matrix $\bbM_1$ resulting from 
    random data. }
  \label{A_randfig}
\end{figure}

For comparison, in Fig.~\ref{A_randfig} we also show an example of a
transition matrix when the CPU usages are replaced with randomly
generated data.  Any reasonable structure is absent here.  These
results are in accord with our intuition that the proposed
probabilistic model for data reduction, Eq.~(\ref{synmod}), although
an incomplete description, still leads to a reasonable reconstruction
of network connectivity.

\section{Modeling gene regulation}

In the example above, as mentioned, data are plentiful. In genomic
time series, data are scarce. However, the above exercise is designed
to test a phenomenological model into which one can incorporate
additional knowledge about genetic regulation. This goal, constraining
possible models by compensating for sparse data with prior knowledge,
is mathematized via Bayesian analysis.

We emphasize that we are {\it not} using Bayesian analysis to attempt
to fit for the innumerable unknowns in a microscopic model, \eg a
chemical kinetics model of transcriptional regulation.
We are not interested in these parameters but in couplings and network
topology. We instead augment the successful phenomenological or
`macroscopic' model above with prior knowledge about transcriptional
regulation.

\subsection{Bayesian statistics}
\label{ss:bayes}

A brief summary of Bayesian statistics is in order.  We refer the
interested reader to standard textbooks (cf.~Ref.~\cite{bayes}) for
discussions of philosophical implications of Bayesian statistics, as
well as standard statistical properties of Bayesian estimators.  We
focus below only on the relevant features.

Bayes rule itself,
\begin{equation}
  P(b|a) =\frac{P(a|b)P(b)}{P(a)} \;,
\end{equation} 
is merely a rewriting of the rules of joint probabilities.  The
connection with interpretation of experiments is made by identifying
$b$ as the model, or specifically the vector of parameters $\btheta$
of the model; and $a$ as the data $D$. Then
\begin{gather}
  P(\btheta|D)=\frac{P(D|\btheta)P(\btheta)}{P(D)}\;,
  \label{posterior}\\
  P(D) = \int d\btheta \, P(\btheta,D)= \int d\btheta\, P(D|\btheta)
  P(\btheta).
\end{gather} 
In Eq.~(\ref{posterior}), the left hand side is called the
{\it{posterior}}, and the first term in the numerator in the right
hand side is called the {\it{likelihood}}. The strength of Bayesian
methods comes from $P(\btheta)$, the {\em prior}.  The prior
summarizes knowledge about the probability of a model {\em before} the
the results, $D$, of an experiment are observed. When experimental
data are abundant the exact specification of the prior is usually
unimportant (\cf Refs.~\cite{clarke-barron-90,nb02}); when data are
scarce the prior constrains the space of available models. While
careless {\it a priori} assumptions may constrain the observer to the
wrong part of the model space, in the analysis of genetic regulatory
networks one may exploit well established knowledge about
transcriptional regulation to construct appropriate priors, as we
illustrate in Sec.~\ref{RegReg}.

Armed with the prior, one next finds the {\it a posteriori} expected
value of the parameters:
\begin{equation}
  \< \btheta \>_D = \int d\btheta\, \btheta P(\btheta|D)
  = \frac{\int d\btheta \,\btheta \, P(\btheta) P(D|\btheta)}{P(D)}
  \equiv  \frac{\< \btheta \,P(D|\btheta) \>}{\< P(D|\btheta)\>}\;,
  \label{avpost}
\end{equation} 
where $\<\dots\>$ and $\<\dots\>_D$ denote expectations over the prior
and the posterior respectively.  When the number of data $N$ is large
we expect the posterior to be tightly peaked around some value
$\widehat{\btheta}$ which maximizes the posterior (maximum a
posteriori probability, or MAP, values), and then $\widehat{\btheta}$
is the first order term in the saddle point asymptotic expansion of
$\<\btheta\>_D$ in powers of $1/N$.

Even for severely undersampled problems, $N$ is usually large enough
so that $\widehat{\btheta} \approx \<\btheta\>_D$, and it is tempting
to replace integrals in Eq.~(\ref{avpost}) by their saddle point
values. One of the greatest realizations in Bayesian theory in the
last decades was that such replacement is wrong
\cite{schwartz,janes,clarke-barron-90,mackay,vijay}. Indeed, when
averaging over all possible models, each of the integrals in
Eq.~(\ref{avpost}) will have a contribution from fluctuations around
the value $\widehat{\btheta}$.  For example, under some very general
conditions, and with an assumption that $\log P(D|\btheta)$ scales
linearly in the number of data, the total probability of the data,
$P(D)$, has the following expansion in powers of $1/N$
\begin{equation}
    \log P(D) =  \log P(D|\widehat{\btheta}) - \frac{K}{2} \log N
     - \frac{1}{2} \log \det \left[\left.\frac{\partial^2}{\partial
          \theta_i \partial \theta_j}\right|_{\widehat{\btheta}}
      \frac{\log P(D|\btheta)}{N}\right] + \log P(\widehat{\btheta})
    +o(N^0)\;,
  \label{occam}
\end{equation}
where $K$ is the number of parameters in the model (dimensionality of
$\btheta$).  The integral in the numerator of Eq.~(\ref{avpost}) can
be written in a similar fashion. We see that the terms beyond the
maximum likelihood contribution $\log P(D|\widehat{\btheta})$ are
generally negative and their magnitude grows with $K$.  Thus these
terms provide a built--in punishment for model complexity.  For this
reason, they are known in the literature as as the Occam razor
(cf.~Ref.~\cite{vijay}).

To illustrate, imagine that the prior admits two model families
$\Theta^1$ and $\Theta^2$ with different parameters $\btheta^1$,
$\btheta^2$, such that $K_1 \equiv \dim \btheta^1 < K_2 \equiv \dim
\btheta^2$.  As an example, consider fitting a function with a
polynomial of low degree ($K_1$) or high degree ($K_2$).  Usually we
would expect the model family with more parameters to be better at
explaining the data: $P(D|\widehat{\btheta^2})>
P(D|\widehat{\btheta^1}).$ Thus if we were to choose a model family
that explains the data best based on the maximum likelihood alone, a
more complex model would win.  However, the estimates within the
larger model family, $\Theta^2$, are less robust to small
fluctuations, and this is picked up by the integration over all
parameters: even though $P(D|\widehat{\btheta^1})$ may be smaller than
$P(D|\widehat{\btheta^2})$, the relation between the probabilities of
the model families $P(\Theta^1|D)$ and $P(\Theta^2|D)$, as determined
by Eqs.~(\ref{avpost},~\ref{occam}), may be different. In particular,
for $N\gg 1$ the likelihood term, which scales linearly with $N$,
always wins, and Bayesian model selection approaches that of the
maximum likelihood.  However, for small $N$ the difference between the
likelihood and the other terms in Eq.~(\ref{occam}) is less profound,
and a simpler model family, which is not the best in explaining the
data, may turn out victorious.

In short, Bayes rule shows how a simpler model may be less likely, yet
more probable.

Before ending our quick review of Bayesian statistics, two more
comments are in order. First, as the model selection arguments are
mostly important for small $N$, where $\log N \sim 1$, it would be a
mistake to ignore $O(1)$ terms in Eq.~(\ref{occam}) and use just $K/2
\log N$ as a model complexity punishment (Schwartz's Bayesian
Information Criterion \cite{schwartz}, also mentioned in
Sec.~\ref{syndatmod}). In particular, we believe that such replacement
may be a cause of a common observation that the Bayesian Criterion
overpunishes complex models (cf.~Ref.~\cite{contextgene}). Second,
even though in this work we will be mostly dealing with finite
parameter models, application of nonparameteric methods to biological
data certainly holds promise.  Occam--type arguments for such cases
have been discussed in, for example, Refs.~\cite{BCS,nb02}.

\subsection{Bayesian inference of regulatory dynamics}
\label{ss:regdyn}

\subsubsection{A simple model}

Let $\bmg_t$ stand for the vector of expressions (mRNA concentrations)
of genes in a microarray experiment at time $t$. The number of genes,
$K_\bmg \equiv \dim \bmg_t$, can be on the order of thousands.
In principle, $\bmg_{t+1}$ can depend on concentrations at all times
that preceded $t+1$. However, if we view the gene expression mechanism
in cells as a realization of some chemical kinetics, governed by first
order differential equations, then it is reasonable to expect that the
concentrations depend only on their immediate past, \ie $\bmg_{t+1}=
{\mathbf f} (\bmg_t)$.\footnote{Support for this choice is strong, as
  such models have been used with great success in the design of small
  genetic networks (see, \eg Ref.~\cite{HastyRev} for a review).
  Further, in a study clustering genes by their dynamics, Ramoni \etal
  tested higher order models, but found that the first order dynamics
  gave the most probable result \cite{Ramoni}. Finally, recall that in
  Sec.~\ref{ss:results} above the first order model turned out the
  most probable, as well.}
We begin with the simplest possible dynamic, a simplified version of
Eqs.~(\ref{synmod}, \ref{noisevar}):
\begin{eqnarray}
  \bmg_{t+1} - \bmg_t &=& \bbM \bmg_t + \bxi_t\;,
  \label{linear}\\
  \<\xi_{i,t}\xi_{j,t'}\>&=& \sigma^2
  \,\delta_{ij}\delta_{t,t'} \;,
  \label{noise}
\end{eqnarray}
where the noise is Gaussian, and $\bbM$ and $\sigma$ are unknown.
This is equivalent to
\begin{equation}
  P(\bmg_{t+1}|\bmg_t, \bbM,\sigma) =
  \frac{1}{\left(2\pi\sigma^2\right)^{K_\bmg/2}}
  \exp \left[-\frac{1}{2\sigma^2}
    \big\<\big| \bmg_{t+1} -\bmg_{t} - \bbM\bmg_t \big|^2\big\>_t\right]\;.
  \label{phenmodel}
\end{equation}
where $\<\cdots\>_t$ indicates empirical averaging over time.

Notice that unlike in Eqs.~(\ref{synmod}, \ref{noisevar}) the noise in
this model is not correlated among the genes; moreover, variances are
equal for all genes.  Below we formulate an extension that
incorporates hidden degrees of freedom (\eg biochemistry) whose
omission may otherwise lead to large or correlated noise
(cf.~Sec.~\ref{ss:hidden}). However, with only a handful of
experiments available we cannot hope that data will be able to
determine millions of elements of a full covariance matrix.  As data
become more plentiful, it may even make sense to bypass
Eqs.~(\ref{linear}, \ref{noise}) completely and pursue model
independent feature extraction (as formulated in Secs.~\ref{modelfree}
and \ref{ss:bottleneck}).  Below we pursue the possibility that the
current simple model exhibits some of the success evidenced in
Sec.~\ref{ss:results}.

\subsubsection{Biological priors}
\label{RegReg}

Even in the simplistic form of Eqs.~(\ref{linear}, \ref{noise}), the
dynamic still contains too many parameters ($\sim K_\bmg^2$) to be
tractable. We therefore search for biologically motivated priors to
constrain the possible values of $\bbM$.  An example of such a prior
would be, for example, that genes with similar regulatory sequences
should be regulated similarly. More directly, genes whose promoters
have similar numbers of certain important motifs should be
co--expressed, an ansatz used with notable success by Bussemaker \etal
\cite{Reduce} in discovering regulatory regions.

This may be expressed in the following prior
\begin{equation} 
  P(\bbM|\mu,\ell) \propto \exp \left[-\frac{\ell^2}{2}
    \sum_{i\neq j,k}^{K_\bmg} \left(\frac{M_{ik}-M_{jk}}{d_{ij}}\right)^2
   -\frac{\mu^2}{2}
   \sum_{ij}^{K_\bmg} M_{ij}^2
 \right]\;,
 \label{prior}
\end{equation}
where the first term punishes for different regulation of genes with
similar regulating sequences, and the second assures proper
normalization of priors by effectively constraining the range of
possible $M_{ij}$.\footnote{Physicists will recognize these as the
  `kinetic' and `mass' terms in a Lagrangian, respectively.}  Here
$d_{ij}$ is a distance function measuring deviation between regulatory
regions of genes $i$ and $j$ in terms of the number of each regulatory
motif appearing, weighted by the relative importance of that motif,
found by considering the entire time series as in Ref.~\cite{Reduce}
or `quality factors' as in Ref.~\cite{MobyDick}.

The parameter $\ell$ plays the role of a smoothing length, as in
Ref.~\cite{BCS}, and, lacking a first--principles estimate of its
value, we must integrate over $\ell$, weighted by an appropriate
prior. As explained in Ref.~\cite{nb02}, it is not only likely that
such integration will choose the proper value of $\ell$ almost
independently of such prior, but it may even balance a slightly
improper choice of the distance measure $d_{ij}$ and the difference
form $(M_{ik}-M_{jk})^2$. The same comments relate to the mass $\mu$
and the noise variance $\sigma$ as well.

An enjoyable feature of this prior is that, like the likelihood,
Eq.~(\ref{phenmodel}), it is exponentially quadratic in the unknowns
$\bbM$. Thus the posterior expectations are Gaussian integrals, which
may be performed analytically using the standard Wigner current
technique \cite{zj-96}.  Following Eq.~(\ref{avpost}) we find for the
{\it a posteriori} values of the connection matrix:
\begin{eqnarray}
  \left< M_{ij}\right>_D &=& \left. \frac{\partial}{\partial
      J_{ij}}\right|_{\bbJ =0}\log  Z(\bbJ) \;,\\
  Z(\bbJ)&=& \int d \sigma \,d\mu \,d \ell P(\sigma) P(\mu) P(\ell) 
  \left[\frac{\det \bcA}{\det (\bcA+\bcG)}\right]^{1/2}
  \nonumber
  \\ 
  &&\times \exp
  \left[ \oh \left(\bbB+\bbJ\right)
      (\bcA+\bcG)^{-1}
      \left(\bbB+\bbJ\right)\right]\;.
  \label{Zj}
\end{eqnarray}
Here the curvature tensor $\bcA$ at the saddle point of
Eq.~(\ref{prior}) is given by
\begin{eqnarray}  
  A_{ij,kl} &=& \left[{\mu^2} + \ell^2 \sum_m^{K_\bmg}
    \left(c_{mj}
      +c_{mi}\right) \right]\delta_{ik}\delta_{jl} -2 \ell^2
  c_{ik}\delta_{jl}\,
\end{eqnarray}
and the time--lagged correlation, equal--time correlation, and
`closeness' (the inverse of the distance matrix) matrices are
\begin{eqnarray}
  \bbB&=&\frac{1}{\sigma^2} \left< 
    (\bmg_{t+1}-\bmg_t)\bmg_t^T\right>_t\\
  \bbG&=&\frac{1}{\sigma^2}\left< \bmg_t\bmg_t^T\right>_t\;,\;\;\;\;
  \bcG= \bbI \otimes \bbG\;,\\ 
  c_{ij} &=& 
  \begin{cases}
    d_{ij}^{-2} & \text{if $i\neq j$},\\
    0& \text{if $i=j$}.
  \end{cases}
\end{eqnarray}

The mass $\mu$ regulates the integrals and is expected to be small.
Then $\det \bcA /\det(\bcA+\bcG)$ scales as a large positive power of
$\ell^2/(\ell^2+{\rm const})$, and therefore decreases as $\ell$
decreases. On the other hand, the exponent in Eq.~(\ref{Zj}) involves
$(\bcA+\bcG)^{-1}$, which scales as $1/(\ell^2 +{\rm const})$ for
large $\ell$. The exponent thus decreases as $\ell$ increases.  We may
then reasonably expect that the integrand in Eq.~(\ref{Zj}) will be
peaked at some non--trivial value of $\ell$; this peak should be sharp
since both $\bbB$ and $\bcG$ involve large number of samples.  This
may be viewed as the smoothing length selection by the data
\cite{nb02}.

Note that the priors over the hyperparameters $\sigma, \mu, \ell$ in
Eq.~(\ref{Zj}) are as yet undefined. Since, as mentioned above (see
also Refs.~\cite{clarke-barron-90}), their actual forms are of little
importance, we may hope that by choosing them appropriately we may be
able to render the integrals in Eq.~(\ref{Zj}) analytically tractable.

\subsubsection{Hidden control}
\label{ss:hidden}

The formalism deserves experimental testing. However, one further
extension offers a substantial improvement.  DNA microarrays offer
only a partial view into the workings of a cell, since numerous
important degrees of freedom remain unobserved.  In mathematical
modeling of the yeast cell cycle, for example, considerable
effort\footnote{See, \eg Ref.~\cite{Tyson} for a review.}  has been
exerted to fine--tune models in which only chemistry controls the
processes, and the genetic expression is a mere passive function of
this chemical control.  While this may or may not prove to be an
accurate characterization, some chemical control unobserved by gene
chip experiments certainly exists. Inclusion will clearly necessitate
a model of some structure other than that of Eq.~(\ref{linear}).
Unlike the assumption of linearity, which merely leads to
misestimation of the values of interactions, this effect may make the
dynamical system appear to be not of first order, or introduce a
gene--dependent noise correlation matrix and artificial couplings
between genes that dominate the real ones. To avoid this, we need to
supplement the vector of genes $\bmg$ with a vector of unknown hidden
degrees of freedom $\bmh$. Then the evolution will take form
\begin{eqnarray}
  \pmbeg \bmg_{t+1} -\bmg_t \\ \bmh_{t+1} - \bmh_t \pmend &=&
  \pmbeg \bbM^{\rm gg} & \bbM^{\rm gh}\\ \bbM^{\rm hg}
  &\bbM^{\rm hh} \pmend
  \pmbeg \bmg_t \\ \bmh_t \pmend +
  \pmbeg \bxi_{t} \\ \bet_{t} \pmend \;,
  \label{linhidden}\\
  \<\xi_{i,t}\xi_{j,t'}\>&=& \sigma^2 \,\delta_{ij}\delta_{t,t'} \;,\\
  \<\eta_t^i\eta_{t'}^j\>&=& 
  \,v_{ij}\delta_{t,t'} \;.
\end{eqnarray} 

Within Bayesian analysis, one may integrate over the unknown degrees
of freedom and their possible couplings while remaining agnostic about
the identities of $\bmh$, and similarly sum over their possible
dimensionality $K_\bmh$. As data are scarce, it is reasonable to
expect that models with small number of hidden units will dominate the
posterior.  Thus we allow a full correlation matrix for the Gaussian
noise in the hidden units, $\bet_t$, since this adds only a few
additional parameters to our model ($\bbM^{\rm gg}$, of course, has a
few million), but allows necessary flexibility.

One will need a prior over newly introduced $\bbM$'s.  Since chemistry
couples to expression via the transcription factors, and therefore via
the regulatory sequences, we can write a similar prior as above,
namely
\begin{equation} 
  P(\bbM^{\rm gh}|\mu_{\rm gh},\ell) \propto \exp \left[-\frac{\ell^2}{2}
    \sum_{i\neq j}^{K_\bmg}\sum_{k}^{K_\bmh} \left(\frac{M^{\rm gh}_{ik}-
        M^{\rm gh}_{jk}}{d_{ij}}\right)^2
    -\frac{\mu_{\rm gh}^2}{2}
    \sum_{i}^{K_\bmg}\sum_j^{K_\bmh} (M^{\rm gh}_{ij})^2
  \right]\;,
  \label{Mgh}
\end{equation}
with a different mass, but the same kinetic term as in
Eq.~(\ref{prior}).  In the absence of precise identification of the
hidden degrees of freedom, we lack a biological prior on $\bbM^{\rm
  hg}, \bbM^{\rm hh}$ and must therefore choose a prior that does not
spoil the analytic tractability of the resulting integrals.


\section{Outlook}
The main difficulty in obtaining time series data is not biological or
technological, but rather financial. Affymetrix chips, the more
reliable of the two dominant technologies, are quite expensive. As
estimated in \cite{rage}, a 24-point time series with replication
factor of 3 currently costs $\sim\$57,600$. As the cost of the
technology decreases, and as data become more plentiful, the role of
priors in inferring connectivity and possible causation diminishes.


Moreover, as noted before, with increasing data comes the possibility
of model independent, nonparametric feature extraction by learning the
joint probability distributions of expression levels at different
times (cf.~Ref.~\cite{BCS}). We highlight two promising such
directions below.

\subsection{Mutual information and entropy distance}
\label{modelfree}
A completely model independent visualization tool of informatics is a
low dimensional embedding of the connectivity diagram of the multiple
genes via some meaningful metric. To this end, armed with a
successfully learned joint probability, one may use information theory
to define distance in a principled way.

An example of such a diagram based on information theoretic ideas is
presented for simulated chemical kinetics in \cite{Arkin01}, in which
the mutual information
\beq I(i\goto j )&=&\int dg_{i} dg^+_j P(g_i,g^+_j) \log_2 \left(
  \frac{P(g_i,g^+_j)}{P(g_i)P(g_j)} \right)
\eeq was used for embedding, with the time--lagged joint probability
of degrees of freedom $P(g_i,g_j^+)\equiv \<P(g_i(t),g_j(t+\tau))\>_t$
learned by histogramming.  While mutual information is a useful
similarity measure, it is not a distance in that it does not obey the
triangle inequality.  However, the `entropy distance' \beq D_H(i\goto
j)=-\int dg_{i} dg^+_j P(g_i,g^+_j)\log_2
\left(\frac{P(g_i,g^+_j)^2\sigma_i\sigma_j}{ P(g_i)P(g_j)}\right) \eeq
does obey the triangle inequality~\cite{MacKay:itp} and thus can be
used as a metric, based on which one can form an embedding in lower
dimensions and construct a process diagram as in
Refs.~\cite{Arkin95,Arkin97,Arkin01}. Here, by $\sigma_{\{i,j\}}$ we
mean the uncertainties in measurements of $g_{\{i,j\}}$, respectively.
Note that this distance is reparameterization invariant
under monotonic reparameterizations $x\goto f(x); y\goto g(y)$,
$\sigma_x\goto \sigma_f/ f'$, etc.

\subsection{Clustering by meaningful information: the information bottleneck}

\label{ss:bottleneck}

Clustering without identifying a specific property of interest is
meaningless. In most cases, if we try to learn from data (fit a curve,
select a model, extrapolate, cluster, etc.) we are doing so not to
find the parameters per se, but to use them to make predictions of
future data \cite{bnt}. Thus in dynamics the variables of relevance
are the future gene expressions, and one should cluster data by
maximally compressing them while retaining the most information about
the subsequent time steps.\footnote{In contrast, genes are usually
  clustered by similarity of their expression levels measured in some
  ad hoc metric \cite{CellCycleGenomics2,contextgene}.}

This idea was put on firm information--theoretic ground recently with
the development of the {\it information bottleneck} \cite{bottleneck},
which, given a joint probability distribution between degrees of
freedom (\eg gene expressions at some time) and a quantity of interest
(\eg gene expressions at subsequent times), allows an iterative
calculation of the meaningful clusters in a probabilistic clustering
algorithm.

An important aspect omitted from current formulations of the
information bottleneck is Bayesian integration over possible joint
probability distributions; this procedure smoothes the data and avoids
clustering noise.  We expect that this will be one of the most
promising as well as principled lines of research in bioinformatics.

\subsection{Prognosis}

The future is promising for such data--driven techniques: data are
becoming more plentiful, computational power continues to
exponentiate, and the data themselves are becoming more reliable, as
those who hope to interpret them study more carefully their statistics
and analyses.\footnote {See, \eg Ref.~\cite{MOM,MOM_dimacs} for one
  such careful analysis of Affymetrix data and Affymetrix's standard
  data analysis.}  However, before any new techniques in computational
biology will be widely exploited, they must be `verified' by comparing
with results agreed upon in the biological community. In this case,
verification will entail corroboration of inferred causal relations
among genes (or within an inferred module of genes) with the
biological literature. We anticipate that such time series based
techniques will find common usage as tests for consensus with known
connectivities become more standardized, and we look forward to their
continued development and implementation.

\subsection*{Acknowledgments}

The authors are grateful to Dimitris A\-na\-stassiou, Guillaume Bal,
William Bia\-lek, Har\-men Busse\-ma\-ker, Michael Elowitz, Stanlslas
Leibler, Chris\-ti\-na Les\-lie, Bud Mish\-ra, Alex Rikun, Burk\-hard
Rost, Ana Ra\-do\-va\-novic, Andrey Rzhetsky, Mi\-sha Samoilov,
Ta\-pi\-o Schnei\-der, Bo\-ris Schrai\-man, Su\-sanne Still, Naftali
Tish\-by, and John Tyson for many stimulating discussions.  The
authors were partially supported by NSF Grant No.\ PHY--9907949 to the
Kav\-li Institute for Theoretical Physics.

\bibliographystyle{abbuns} \bibliography{regulation}
\end{document}